\documentclass[aps,prl,preprint,groupedaddress]{revtex4-1}

\usepackage[usenames,dvipsnames,svgnames,table]{xcolor}
\usepackage[]{graphicx}
\usepackage[multidot]{grffile}
\usepackage{booktabs}

\usepackage{amsmath}
\usepackage{ams}

\newcommand{\E}{\mathcal{E}}
\newcommand{\lb}{\left(}
\newcommand{\rb}{\right)}
\newcommand{\dpsi}{\delta\psi}

\renewcommand\Re{\operatorname{Re}}
\renewcommand\Im{\operatorname{Im}}

\begin{document}


\title{Timing jitter of passively mode-locked semiconductor lasers subject to optical feedback; a semi-analytic approach}


\author{Lina Jaurigue\textsuperscript{a}, Alexander Pimenov\textsuperscript{c},  Dmitrii Rachinskii\textsuperscript{d}, Eckehard Sch{\"o}ll\textsuperscript{a}, Kathy L{\"u}dge\textsuperscript{b}, Andrei G. Vladimirov\textsuperscript{c,e}}

\affiliation{(a) Institut f{\"u}r Theoretische Physik, Sekr. EW 7-1, Technische
Universit{\"a}t Berlin, Hardenbergstr. 36, 10623 Berlin, Germany}

\affiliation{(b) Department of Physics, Freie
Universit{\"a}t Berlin, Arnimallee 14, 14195 Berlin, Germany}

\affiliation{(c) Weierstrass Institute, Mohrenstr. 39, 10117 Berlin, Germany}

\affiliation{(d) Department of Mathematical Sciences, The University of Texas at Dallas, 800 W. Campbell Road, Richardson, Texas 75080, USA}

\affiliation{(e) Lobachevsky University of Nizhny Novgorod, Russia}


\date{\today}

\begin{abstract}
We propose a semi-analytical method of calculating the  timing fluctuations in mode-locked
semiconductor lasers and apply it to study the effect of delayed coherent optical feedback on pulse
timing jitter in these lasers. The proposed method greatly reduces computation times and therefore
allows for the investigation of the dependence of timing fluctuations over greater parameter
domains. We show that resonant feedback leads to a reduction in the timing jitter and that a
frequency-pulling region forms about the main resonances, within which a timing jitter
reduction is observed. The width of these frequency-pulling regions increases linearly with short
feedback delay times. We derive an analytic expression for the timing
jitter, which predicts a monotonous decrease in the timing jitter for resonant feedback of
increasing delay lengths, when timing jitter effects are fully separated from amplitude jitter
effects. For long feedback cavities the decrease in timing jitter scales  approximately as $1/\tau$
with the increase of the feedback delay time $\tau$.
\end{abstract}

\pacs{}

\maketitle

\section{}
\subsection{}
\subsubsection{}

\section{Introduction}

Many current and future applications require ultra-high repetition frequency light pulse
sources \cite{RAF07}. Among these applications most also require highly regular pulse arrival times.
Mode-locked (ML) solid state lasers can fulfill these requirements.
However, such devices are too expensive for large scale use. Due to this limitation extensive
research has gone into semiconductor ML lasers. The most attractive mode-locking technique, due to
its simplicity of production and handling, is passive mode-locking, which does not require any
external RF modulation source. However, due to the absence of an external reference
clock passively ML lasers exhibit relatively large fluctuations in the temporal positions
of pulses compared with a perfectly periodic pulse train \cite{LIN10c}. This phenomenon is referred
to as pulse timing jitter. Recently, it was proposed to use optical feedback to significantly reduce
the  timing jitter of passively ML lasers \cite{SOL93,LIN10e,OTT12a,OTT14b}. Other methods of pulse
stream stabilisation which have been investigated include hybrid mode-locking \cite{FIO10,ARK13} and optical
injection \cite{REB10,REB11}.
To characterize the performance of such devices, with respect to the timing regularity, the timing
jitter is calculated. Experimentally this is done using the von Linde method, which involves
integrating over the sidebands of the power spectrum of the laser output. However, for the
numerical investigation of ML lasers the von Linde method can be impractical as it is computationally
very expensive. In this paper we therefore propose a semi-analytical method of calculating the pulse
timing jitter for a set of delay differential equations (DDEs) proposed earlier to describe passive
mode-locking in semiconductor lasers \cite{VLA04,VLA04a,VLA05}. The method is of general nature and can be used to estimate the
variance of timing fluctuations in a wide range of time periodic dynamical systems described by
autonomous systems of DDEs subject to weak additive noise.

Theoretical analysis of the influence of noise on ML pulses propagating in a laser cavity was first
performed by H. Haus using a master equation \cite{HAU93a}.  Later this technique was extended by
taking into account the finite carrier density relaxation rate in semiconductor lasers \cite{JIA01}. The
master equation has secant-shaped ML pulses as a solution, and a small perturbation of
this state can be studied using the linearized equation of motion. The perturbed pulse is described
by four parameters: the perturbations of the pulse amplitude, phase, frequency, and timing. Using
the orthogonality of the solutions of the linearized equation to the solutions of the adjoint
homogeneous linear system, coupled first order differential equations of motion, driven by noise,
can be written out.  However, due to multiple simplifying assumptions underlying the Haus master
equation, this approach is not directly applicable to the analysis of semiconductor laser devices.
This is why the theoretical estimation
of timing jitter in ML semiconductor lasers has been previously performed using the direct numerical
simulations of travelling wave \cite{ZHU97,MUL06} and delay-differential equation (DDE)
\cite{OTT12a,OTT14b,JAU15a,SIM14} models. As purely computational approaches are time-consuming, the influence of
noise on the dynamics of ML pulses has been studied only in limited parameter regions. In a recent
paper \cite{PIM14b} a new semi-analytical method to estimate timing jitter in the DDE-model
\cite{VLA04,VLA04a,VLA05} of a passively ML semiconductor laser  was proposed. This method was used
to study the effect of nonlinear
phenomena such as bifurcations and bistability on timing jitter, and the numerical results were found to be in
good qualitative agreement with experimental data. In this paper we consider a generalisation of
the semi-analytical method to study passively ML lasers
with multiple delayed feedback.  We then use this semi-analytical method to
derive a formula for the timing jitter for resonant feedback delay lengths.

In Section~II \ref{sec:method} we introduce an autonomous DDE model of a laser
operating in a passive ML regime and describe the parameters used in our calculations.
In Sec.~III, by linearizing the model equations near the ML periodic solution and projecting the
perturbation term on the neutral eigenfunctions corresponding to the time and phase shift
symmetries of the unperturbed equations, we derive a semi-analytical expression for the variance of the
pulse timing fluctuations \cite{HAL66,HAL77}.
Section~IV is devoted to the comparison of the results obtained using this expression with those
of direct numerical calculations of pulse timing jitter, and a derivation of the dependence of the
timing jitter on the feedback delay time in the particular case of resonant feedback. Finally, in
Sec.~V we conclude with a brief discussion of our results.
\cite{FLU07}

\section{DDE Model}

\begin{figure}[t!]
\begin{center}
\includegraphics[width=0.49\textwidth]{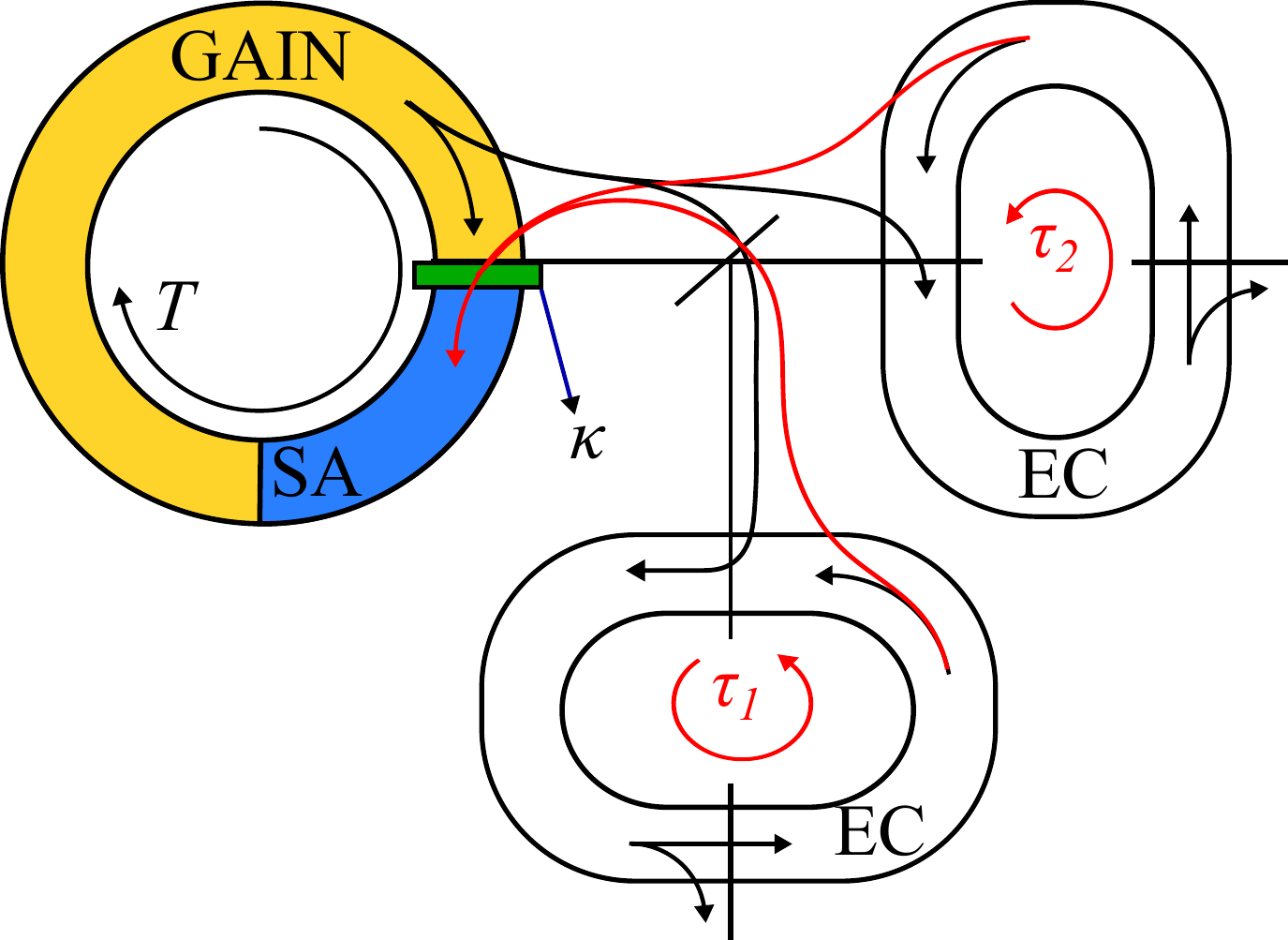}
\caption{Schematic diagram of a two section ring cavity laser subject to optical feedback from two external cavities (EC). The yellow region represents the gain section, the blue region corresponds to the saturable absorber (SA) section and the green bar indicates the spectral filtering element.
}\label{model}
\end{center}
\end{figure}

We use a DDE model for a passively ML ring cavity laser subject to optical feedback from $M$
external cavities,
based on the model introduced in \cite{OTT12a}, a schematic diagram of the model is shown in Fig.~\ref{model} for the case of
two feedback cavities.
This model is an extension of the DDE model proposed in \cite{VLA04,VLA05}. A detailed description
and derivation of the
feedback terms for a laser with a single feedback cavity can be found in \cite{OTT12a}. The final
set of three coupled delay differential equations is
\begin{eqnarray}
& \qquad \dot{\mathcal{E}}\left(t\right)=-\lb \gamma+i\omega\rb\mathcal{E}\left(t\right)
+\gamma R\left(t-T\right) e^{-i \lb\Delta \Omega+\omega\rb T}
\mathcal{E}\left(t-T\right)\nonumber\\
 &+\gamma\sum_{m=1}^M\sum^{\infty}_{l=1} K_{m,l} e^{-i l C_m} R\left(t-T-l\tau_m\right) e^{-i\lb \Delta \Omega+\omega \rb \left(T+l\tau_m\right)}
 \mathcal{E}\left(t-T-l\tau_m\right)+D\xi\left(t\right), \label{subeq:E}\\
 &\dot{G}\left(t\right)= J_g-\gamma_g G\left(t\right)-e^{-Q\left(t\right)}\left(e^{G\left(t\right)}-1\right)|\mathcal{E}\left(t\right)|^2,\label{subeq:G}\\
 &\dot{Q}\left(t\right)=J_q-\gamma_q Q\left(t\right)-r_s e^{-Q\left(t\right)}\left(e^{Q\left(t\right)}-1\right)|\mathcal{E}\left(t\right)|^2,\label{subeq:Q}
\end{eqnarray}
with
\begin{equation}
 R\left(t\right)\equiv\sqrt{\kappa} e^{\frac{1}{2}\left(\left(1-i \alpha_g\right)G\left(t\right)-\left(1-i \alpha_q\right)Q\left(t\right)\right)}.\label{defR}
\end{equation}
The dynamical variables are the slowly varying electric field amplitude $\mathcal{E}$, the saturable
gain $G$ and the saturable loss $Q$. The saturable gain $G$ and saturable loss $Q$ are related to
the carrier inversion in the gain and absorber sections, respectively.  In Eq.~\eqref{subeq:G} $J_g$
is related to the current pumped into the gain section and $J_q$ in Eq.~\eqref{subeq:Q} describes
the unsaturated absorption. The carrier lifetimes in the gain and absorber sections are given by
$1/\gamma_g$ and $1/\gamma_q$, respectively.  The factor $r_s$ is the ratio of the saturation
intensities in the gain and absorber sections. The $M+1$ delay times in this system are the cold
cavity round-trip time $T$ and the external cavity round-trip times (delay times) $\tau_m$ of the
$M$ feedback cavites. The cold cavity round trip time is defined as $T$\,$\equiv$\,$v/L$, where $L$
is the length of the ring cavity. The bandwidth of the laser is limited by the finite width of the
gain spectrum, which is taken into account by a Lorentzian-shaped filter function of width
$\gamma$.  $\omega$ describes the shift between the reference frequency and the central
frequency of the spectral filter. The possibility of detuning between this latter frequency and the
frequency of the nearest cavity mode is allowed for by the inclusion of $\Delta \Omega$. The optical
feedback is described by the sum in Eq.~\eqref{subeq:E}. Here $l$ is the number of round-trips in
the external cavity, $K_{m,l}$ is the round-trip dependent feedback strength of the $m$th feedback
cavity and $C_m$ is the phase shift that accumulates over one round-trip in the external
cavity. Below we consider feedback contributions only from light that has made one round-trip in the
external cavities ($K_{m,1}=K_m$). The last term in Eq.~\eqref{subeq:E} models spontaneous emission
noise using a complex Gaussian white noise term $\xi(t) = \xi_1(t) + i \xi_2(t)$ with strength $D$,
$$ \left\langle \xi_i(t)\right\rangle = 0\qquad \mbox{and}
\qquad\left\langle\xi_i(t)\xi_j(t')\right\rangle=\delta_{i,j}\delta(t-t').$$ Equation~\eqref{defR}
describes the amplification and losses of the electric field during one round-trip in the laser
cavity.  Internal and out-coupling losses are taken into account in the attenuation factor $\kappa$
and the linewidth enhancement factors ($\alpha$-factor) in the gain and absorber sections are denoted
$\alpha_g$ and $\alpha_q$, respectively.

\begin{table}[htbp]
  \centering
  \begin{tabular}[c]{l l l l l} 
    \hline
    symbol & value && symbol & value\\
    \toprule\hline
     $T$ & $25$ ps &&$\gamma$ &  $2.66$ ps$^{-1}$\\
     $\gamma_g$ & $1$ ns$^{-1}$ &&$\gamma_q$ & $75$ ns$^{-1}$\\
    $J_g$ & $0.12$ ps$^{-1}$&&$J_q$ & $0.3$ ps$^{-1}$\\
    $r_{s}$ & $25.0$  && $C_m$ & $0$ \\
    $\kappa$ & $0.1$ &&$\Delta\Omega$& 0 \\
    \bottomrule
  \end{tabular}
  \caption{Parameter values used in numerical simulations, unless stated otherwise.}
  \label{tab:Simparams}
\end{table}

\section{Perturbation analysis}
\label{sec:method}

Various methods of calculating the timing jitter are discussed in
\cite{MUL06,OTT14b,KEF08,PIM14b,LIN86}. In this section, we consider an extension of the
semi-analytical method of timing jitter estimation proposed in \cite{PIM14b} for the DDE model of
passively
ML laser to the system \eqref{subeq:E}-\eqref{subeq:Q} with external feedback and, hence, multiple
delay times. Details of the derivation of the semi-analytical expression for the
estimation of pulse timing jitter are presented in the Appendix. As we do not use a specific form of equations \eqref{subeq:E}-\eqref{subeq:Q},
the same approach can be applied to the analysis of the effect of small additive noise on stable periodic
solutions in other physical systems described by autonomous DDEs with multiple delays.
 The advantage of the proposed method, compared with the von Linde technique or the so called long term jitter
calculation \cite{OTT14b}, is that it is based on the numerical solution of deterministic equations and
therefore requires much shorter computation times. Furthermore, when the spontaneous emission noise
is modeled by a Gaussian white noise term,
the fluctuations of the pulse arrival times behave like a random walk \cite{OTT14b}, making the
timing jitter calculated from the semi-analytical method proportional to the rms timing jitter given
by the von Linde method. This is useful for comparison with experiments.

We consider a periodic ML solution, $\psi_0 = (\Re \mathcal{E}_0, \Im \mathcal{E}_0, G_0, Q_0)^{\mathrm{T}}$
of the system \eqref{subeq:E}-\eqref{subeq:Q} for $D = 0$, with period $T_0$. One should note that due to the rotational symmetry,
there is a family of such solutions $\Gamma_\varphi\cdot \psi_0 = (\Re (e^{i\varphi}\mathcal{E}_0), \Im (e^{i\varphi}\mathcal{E}_0), G_0, Q_0)^{\mathrm{T}}$,
where $\Gamma_\varphi$ denotes the corresponding matrix of rotation of the $\mathcal{E}_0$ plane.
The noise perturbation is assumed to be reasonably small, $D\ll 1$,
and we restrict our analysis to the situation when solutions remain at a distance of order $D$
from the torus of stable periodic solutions $\Gamma_\varphi\cdot \psi_0(t+\theta)$  at all times (that is, the probability of a large fluctuation of the solution
is assumed to be negligible during the typical time interval of system observation).
Under this assumption, the noise results in a slow diffusion of the phase $\theta$ of the solution, as well as a slow diffusion of the angular variable $\varphi$.
Furthermore, one expects that the variance of the phase $\theta$ and of the variable $\varphi$ increases linearly with
time, that is $\langle \theta -\bar \theta\rangle^2 \propto t$, which expresses a simple
diffusion process \cite{Daffertshofer}. We use the coefficient of proportionality
in this relationship as a measure of the timing jitter.

The phase of a solution can be defined in several ways \cite{rice}, which, in practice, lead
to equivalent or close results when applied for the evaluation of the phase diffusion rate.
In particular, the definition of the asymptotic phase is based on the fact that every solution $\psi(t)$
of the unperturbed system \eqref{subeq:E}-\eqref{subeq:Q} with $D = 0$ converges to
a periodic solution $\Gamma_\varphi \cdot \psi_0(t+\theta)$ in the limit $t\to\infty$
where the constant $\theta$, called the asymptotic phase, and the angle $\varphi$ are specific
to the initial state of the solution $\psi(t)$.
Recall that states of system \eqref{subeq:E}-\eqref{subeq:Q} are functions defined on the interval $[-\tau_M',0]$.
The asymptotic phase $\theta$ and the angle $\varphi$ remain constant along the trajectories of the unperturbed system.
However, in the perturbed system, the asymptotic phase $\theta$ and the angular variable $\varphi$ evolve as functions of the evolving state $\psi(t+r)$ ($r\in [-\tau_M',0]$).

As dynamics are restricted to a small neighborhood of the limit cycle $\psi_0$ (and its rotations $\Gamma_\varphi \cdot \psi_0$), the evolution of the phase can be deduced
from the linearization \eqref{linsys} of system  Eqs.\,\eqref{subeq:E}-\eqref{subeq:Q}
around this cycle. Details on the analysis of the dynamics of the solutions of the linear system \eqref{linsys} as well as its effect on the evolution of the phase can be found in Appendix. 
Noise results in a slow diffusion of the variables $\theta$ and $\varphi$ along
the neutral periodic eigenmodes of the linearized unperturbed system \eqref{eq:linearh} with the variance proportional to time.
There are two such neutral modes,
\begin{equation}\label{eq:neutral}
 \delta
\psi_{1}(t)=
(\Re\dot{\mathcal{E}}_0(t), \Im \dot {\mathcal{E}}_0(t),\dot G_0(t),\dot Q_0(t))^{\mathrm{T}},\qquad
\delta \psi_{2}(t)= (-\Im \mathcal{E}_0(t), \Re \mathcal{E}_0(t),0,0)^{\mathrm{T}},
\end{equation}
which correspond to the time-shift and rotational symmetries of the unperturbed ($D=0$) nonlinear system
\eqref{subeq:E}-\eqref{subeq:Q}, respectively; all the other Floquet modes are exponentially decaying.
Two properly normalized \eqref{eq:orth} neutral modes $\delta \psi_{1}^\dagger(t)$ and $\delta \psi_{2}^\dagger(t)$  of the adjoint linear system \eqref{eq:linearadj} can be used for
calculating the projections of noise onto the eigendirections $\delta \psi_{1}$ and $\delta \psi_{2}$.
Using the perturbation expansion with respect to the small parameter $D$, and adapting the asymptotic analysis from \cite{REB11}, we obtain the following equations
for the noise-driven slow evolution of the phase $\theta$ and the angular variable $\varphi$ of solutions to Eqs.\,\eqref{subeq:E}-\eqref{subeq:Q}:
\begin{equation}\label{rate}
\dot \theta = D\,\delta \psi_{1}^\dagger(t+\theta) \Gamma_{-\varphi} w(t),\qquad \dot \varphi =D\,\delta \psi_{2}^\dagger(t+\theta) \Gamma_{-\varphi} w(t)
\end{equation}
with the Langevin term $\Gamma_{-\varphi}w(t)=(\xi_1(t)\cos\varphi+\xi_2(t)\sin\varphi,-\xi_1(t)\sin\varphi+\xi_2(t)\cos\varphi,0,0)^{\rm T}$
and the $T_0$-periodic coefficients $\delta\psi_{1}^\dagger$ and $\delta \psi_{2}^\dagger$.

The coefficients of the Fokker-Planck equation for the joint probability density
$p(t,\theta,\varphi)$ of the stochastic process \eqref{rate} are also periodic with respect to time.
Since, for $D\ll 1$, the probability density function $p(t,\theta,\varphi)$ changes slowly, Eqs.~\eqref{rate} and the corresponding
Fokker-Planck equation can be averaged over the period $T_0$ of the functions $\delta \psi_{i}^\dagger(t+\theta)$,
resulting in the diffusion equation with constant coefficients \cite{averaging}. The diffusion coefficient 
\begin{equation}\label{dbar}
\bar d_{11}= \frac{D^2}{T_0} \int_0^{T_0}  \bigl(\delta \psi_{1, 1}^\dagger (s) \bigr)^2+ \bigl(\delta \psi_{1, 2}^\dagger(s)\bigr)^2\, ds
\end{equation}
of the time averaged Fokker-Planck equation approximates the rate of diffusion of the phase $\theta$ (see Appendix). 
Finally, since the pulse timing jitter is usually calculated over a long time interval $n \tilde T_0$ with $n \gg 1$ and the average period $\tilde T_0 \approx T_0$,
and is normalized by the number of round-trips $n$, we make the estimate of timing jitter as the product of the diffusion rate by the period
\begin{equation}\label{eq:jitter}
\sigma_{\mbox{var}}^2 = \bar d_{11} T_0 = D^2 \int_0^{T_0} \bigl(\delta \psi_{1, 1}^\dagger (s) \bigr)^2+ \bigl(\delta \psi_{1, 2}^\dagger(s)\bigr)^2\, ds.
\end{equation}
This value is approximately equal to the variance of $\theta(n \tilde T_0)$ divided by $n \gg 1$. We note that for the number of roundtrips $n \geq 1$ that is not sufficiently large,
the numerically calculated timing jitter is not
approximated by \eqref{eq:jitter} since the numerically calculated value is affected by amplitude noise, or, in other words, stable eigendirections play role as well (see Fig. \ref{fig1_1} (a)).

For the case of resonant optical feedback, expression \eqref{eq:jitter} for the timing jitter can be further simplified, to ascertain the dependence on the feedback delay
length. This will be shown in the next section where we compare the analytic result with a numerical estimate of the timing jitter.

\section{Results}
\subsection{Comparison of semi-analytical and numerical methods of timing jitter calculation.}

In this section we compare the timing jitter calculated using Eq.~\eqref{eq:jitter} with that
obtained from the variance of the pulse timing fluctuations (long-term timing jitter) through numerical integration of the
stochastic system (Eqs.~\eqref{subeq:E}-\eqref{subeq:Q} with $D\neq 0$).
The latter (numerical) method is described in detail in \cite{OTT14b}. We will focus mainly on the
case of one feedback cavity, $M=1$, and compare the two approaches to the timing jitter calculation
at different feedback delay times ($\tau_1\equiv\tau$) and the feedback strengths ($K_1\equiv K$).

\begin{figure}[t!]
\centering{}%
\includegraphics[clip,width=0.45\textwidth]{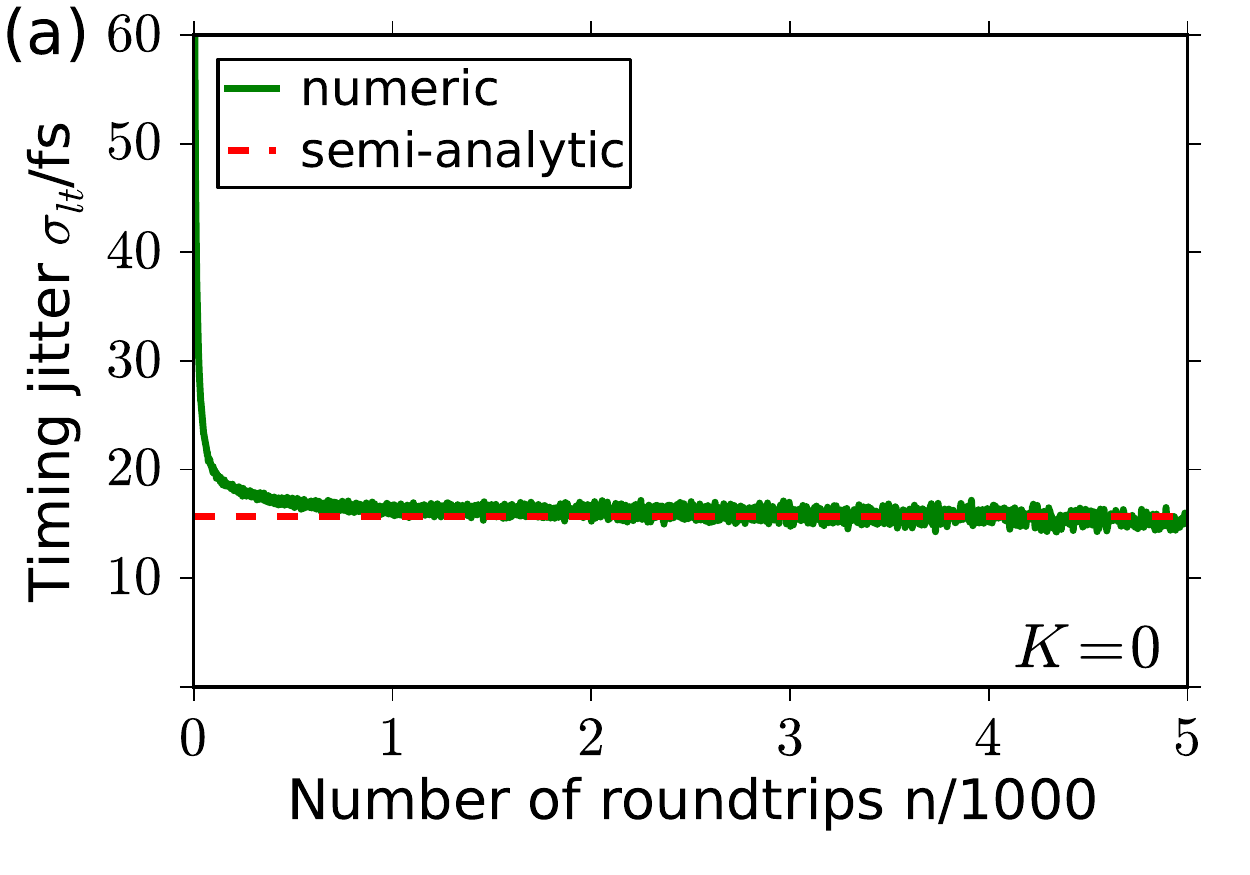}
\label{fig1_1a}
\includegraphics[clip,width=0.45\textwidth]{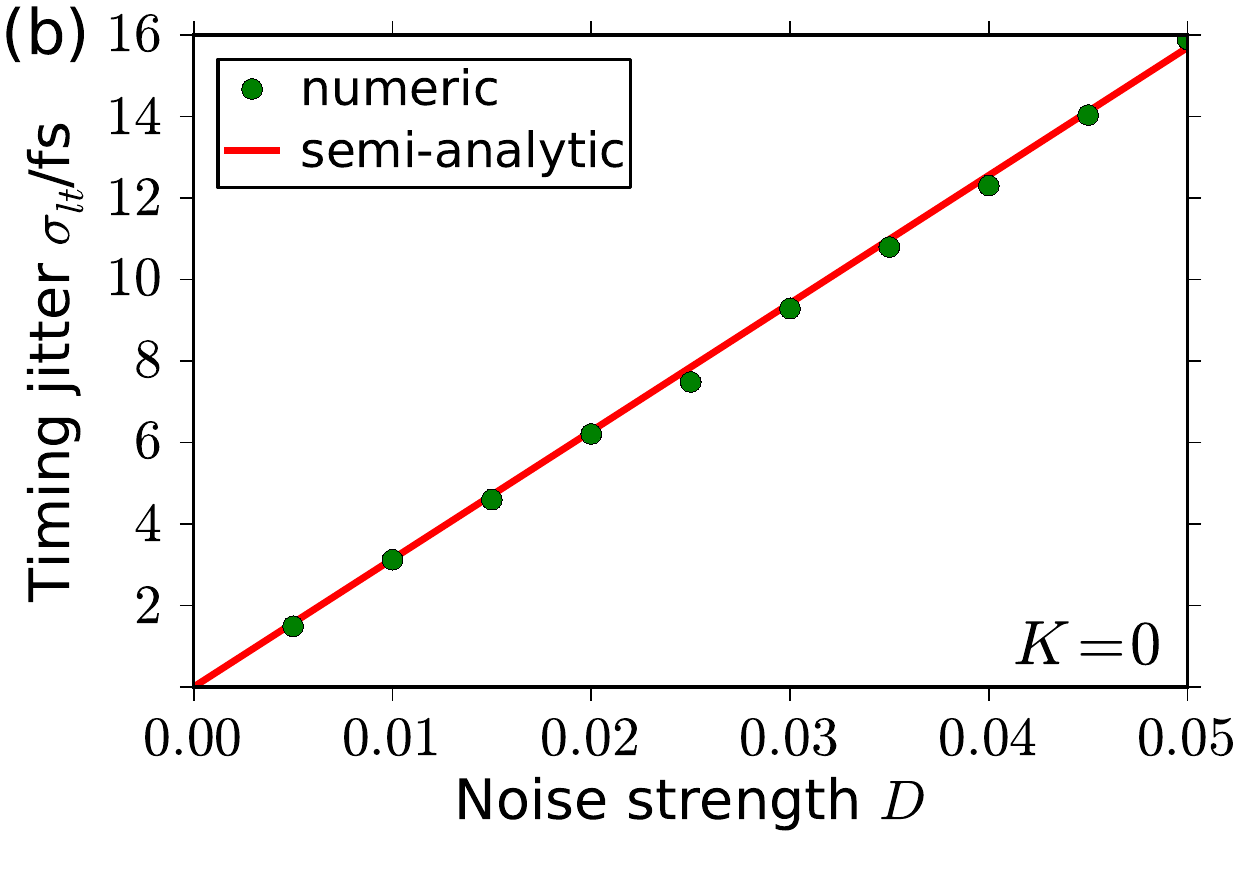}
\label{fig1_1b}

\caption{(a) Comparison of the results of numerical calculation of pulse timing jitter (green solid line),
obtained for different numbers of round-trips $n$, with the timing jitter value from formula
\eqref{eq:jitter} (red dashed line).
(b) Estimation of timing jitter, calculated using formula \eqref{eq:jitter} (red solid line) and the
numerical method (green dots), vs noise strength $D$.
Parameters: $K=0$, $\tau=0$, $T= 25\mbox{ ps}$, $\kappa=0.3$, $\gamma^{-1}=125\mbox{ fs}$, $\gamma_g^{-1}=500\mbox{
ps}$,
$\gamma_q^{-1}=5\mbox{ ps}$, $s=10$, $q_0^{-1}=10\mbox{ ps}$, $g_0^{-1}=250\mbox{ ps}$,
$\alpha_g=2$, $\alpha_q=1$.}
\label{fig1_1}
\end{figure}

First, we apply the semi-analytical method of the timing jitter calculation to the case of a
passively ML semiconductor laser without feedback, i.e. $K_m \equiv 0$ in
Eqs.\,\eqref{subeq:E}-\eqref{subeq:Q}.
In \cite{OTT14b} it was shown that after a sufficiently large number of roundtrips $n$ within the laser
cavity the variance of the pulse timing fluctuations grows linearly with the round-trip number.
In the numerical method the timing fluctuations are therefore calculated over many thousands of cavity
roundtrips. In Fig.~\ref{fig1_1} (a) the timing jitter is plotted as a function of the round
trip number $n$. The initial decrease of the numerically calculated timing fluctuation variance (green line) with
$n$ (for small $n$) can be attributed to the impact of the eigenfunctions with $\Re \lambda <0$ (see Appendix).
Using DDE-BIFTOOL \cite{ENG01}, for $\gamma T \gg 1$ (or $\gamma \tau_m\gg1$), one can typically observe that many characteristic exponents
$\lambda$ of the ML solution have real parts close to $0$, and, therefore, the equation of motion
\eqref{eq:motion} suggests that such exponents will have a non-negligible impact on the numerically
calculated timing  jitter even after many cavity round-trips. Since the eigenfunctions with $\Re
\lambda <0$ are neglected in the semi-analytical approach, the value of the timing jitter estimated using this approach does not depend on
$n$ (dashed red line in Fig.~\ref{fig1_1} (a)). In the limit of large $n$ this value is in agreement with the data obtained by direct numerical
integration of Eqs.\,\eqref{subeq:E}-\eqref{subeq:Q}, as shown in Fig.~\ref{fig1_1} (a).
Figure \ref{fig1_1} (b) shows the timing jitter, obtained using both methods, in dependence of the noise strength $D$. It is seen
that good quantitative agreement is obtained for small to moderate levels of noise.

\begin{figure}[h]
\begin{center}
\includegraphics[width=0.45\textwidth]{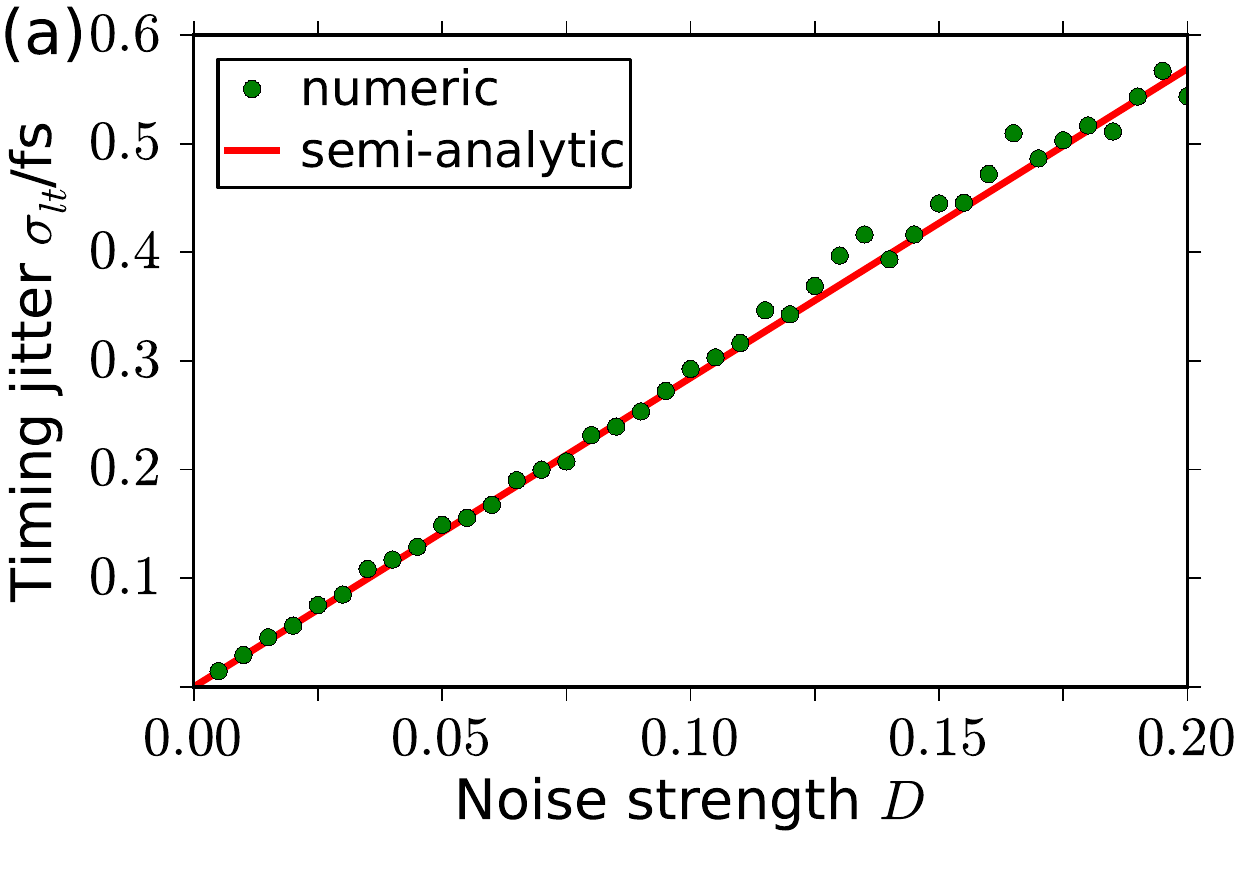}
\includegraphics[width=0.45\textwidth]{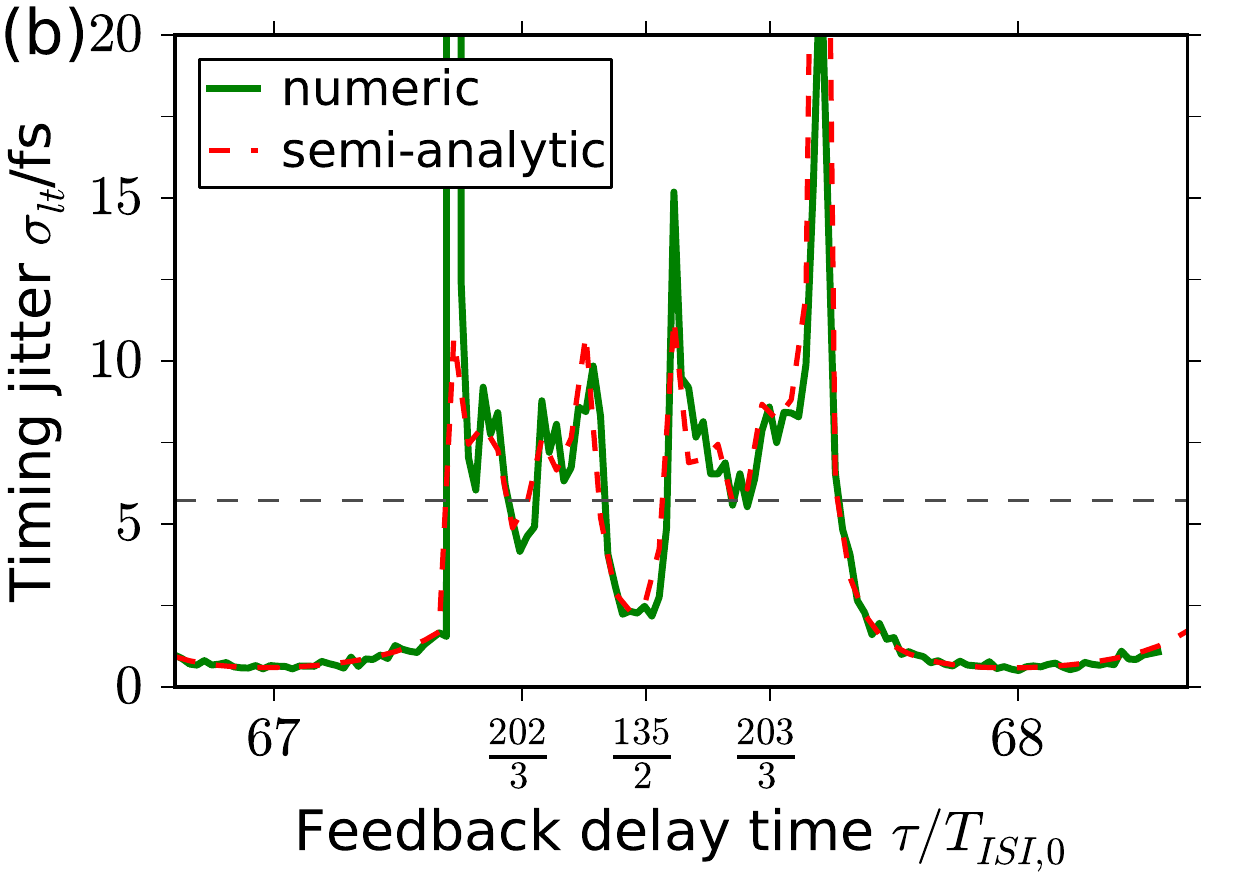}
\caption{(a) Timing jitter in dependence of the noise strength, calculated using the semi-analytic method
(red line) and the numerical method described in \cite{OTT14b} (green dots) for $\tau=70T_{ISI,0}$.
(b) Timing jitter in dependence of the feedback delay time,
calculated using the semi-analytic method (red dashed line) and the numerical method (green line) for $D=0.2$.
Parameters: $\alpha_g=0$, $\alpha_q=0$, $K=0.1$. Other parameters are as in Table \ref{tab:Simparams}.
}\label{pl2}
\end{center}
\end{figure}
Next, let us consider a system with feedback from one external cavity. Figure \ref{pl2} (a)
shows a comparison of the timing jitter calculated from the two methods in dependence of the noise
strength. For the numerical timing jitter calculation method (green dots) the timing fluctuations that
arise over $40000$ round-trips in the laser cavity are calculated, and the variance of these timing
fluctuations is then calculated for $300$ noise realisations. For the semi-analytical method (red
line) the solutions to the adjoint linearized homogeneous system~\eqref{eq:linearadj} are
numerically calculated. In both cases we simulate for a sufficiently long time (approximately 5000 roundtrips) before
starting the calculation of the timing jitter to avoid transient effects. We find very good
agreement between the results obtained using the two methods. For the simulations presented in
Fig.~\ref{pl2} (a) the feedback delay time was chosen to be resonant with the  ML pulse repetition
period (inter-spike interval time) $T_{ISI,0}$ of a solitary laser (ML laser without feedback),
meaning that the condition $\tau=qT_{ISI,0}$ is fulfilled, where $q$ is an integer.
Resonant feedback applied in the fundamental ML regime does not significantly affect the dynamical behaviour of the system,
hence the laser output remains periodic and the semi-analytic method is applicable.
When the feedback delay time is tuned from one resonance to the next, bifurcations
can occur and the dynamical behaviour can change. This is described in detail in \cite{OTT14} and
\cite{OTT12a}. In Fig.~\ref{pl2} (b) the numerically calculated dependence of the timing jitter on
the delay time $\tau$ is compared to that estimated semi-analytically, spanning from the $67$th to
the $68$th resonance ($q=67$ and $q=68$, respectively). Within the frequency-pulling
regions of the main resonances there is very good agreement between the results obtained using the
two methods. The frequency pulling regions are the $\tau$ ranges about the main resonances within which there is
one pulse in the laser cavity and the repetition rate tunes with $\tau$ \cite{OTT12a}. In Fig.~\ref{pl2} (b) these
regions can be identified by the low timing jitter about the main resonances.
At the edges of the frequency-pulling regions there is a sharp increase in the timing jitter. This very
large timing jitter coincides with saddle-node bifurcation points of the deterministic system
(Eqs.~\eqref{subeq:E}-\eqref{subeq:Q} with $D=0$) \cite{OTT14b}. At the edge of the $67$th resonance
there is a large discrepancy between the semi-analytical and numerical methods. This is because in the stochastic system
noise induced switching between bistable solutions,
which arise due to the saddle-node bifurcations, occurs. Away from the bifurcation points there is good agreement between the two methods, also
between the main resonances, because although the dynamical behaviour changes between the main resonances, i.e. multiple feedback induced
pulses, the solutions remain periodic and therefore the semi-analytical method is applicable.

\begin{figure}[t!]
\begin{center}
\includegraphics[width=0.48\textwidth]{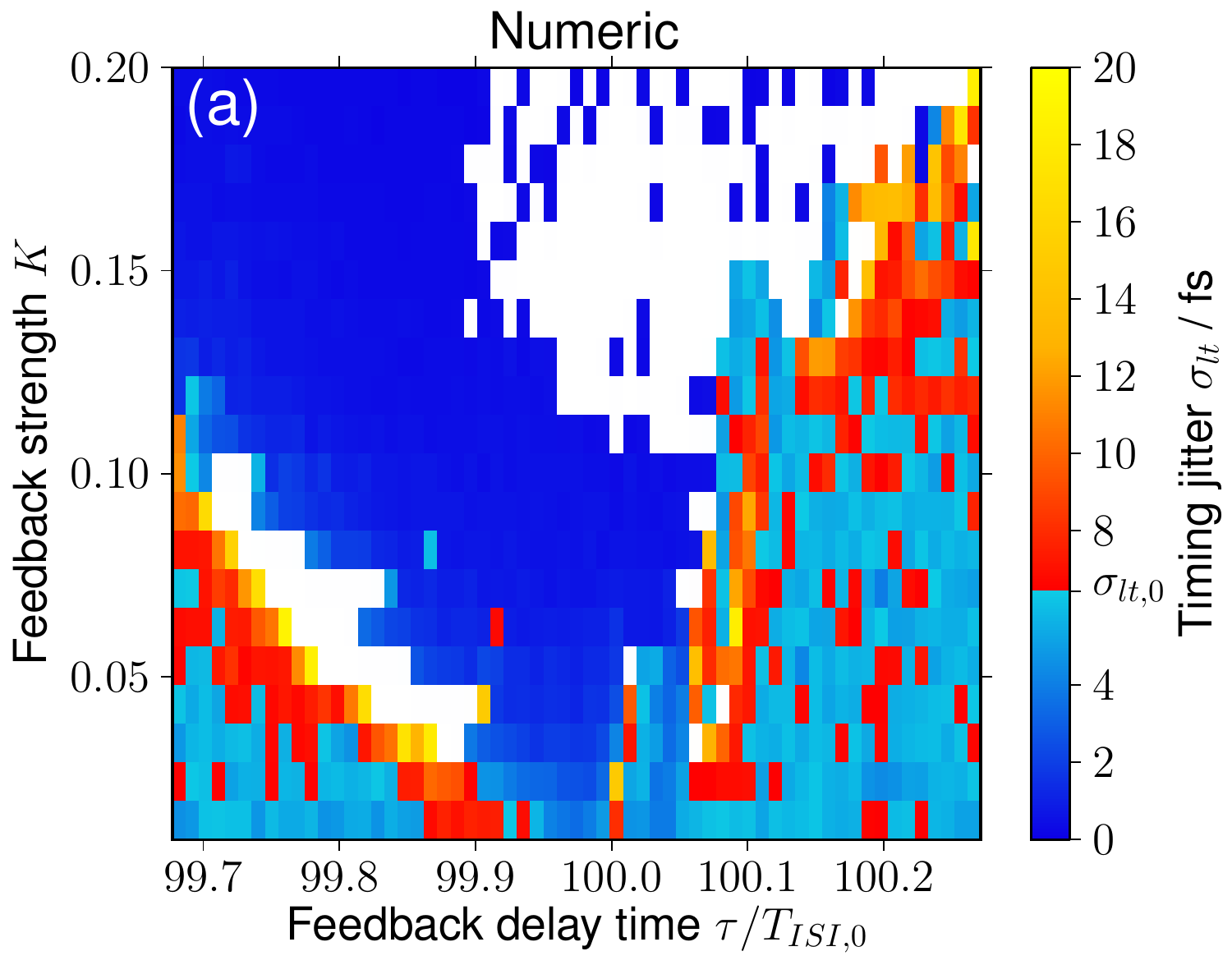}  \includegraphics[width=0.48\textwidth]{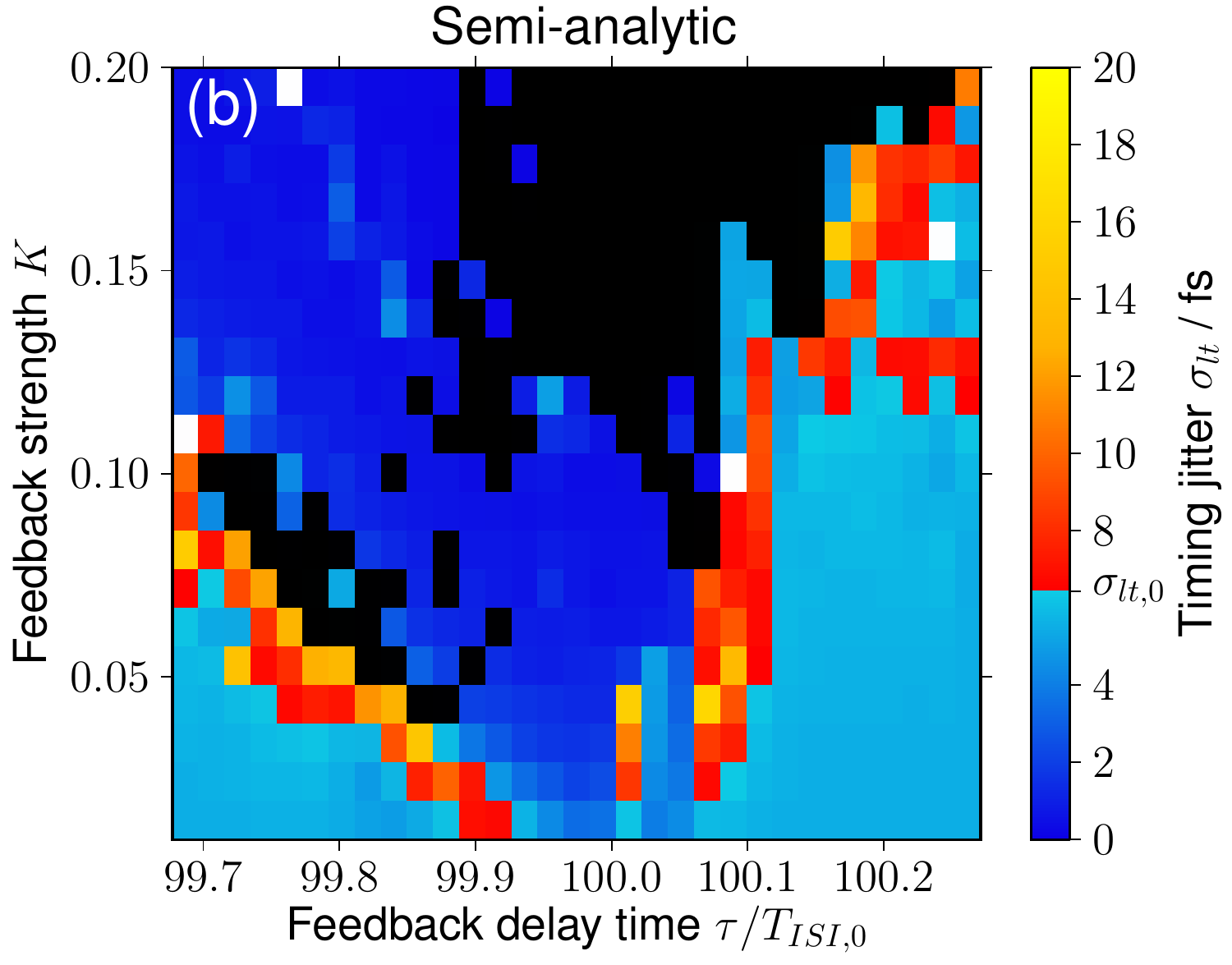}
\caption{Timing jitter in dependence of the feedback cavity delay time and feedback strength,
calculated numerically (a) and using the semi-analytical method (b). The timing jitter is indicated by the colour code and $\sigma_{lt,0}$ is the
timing jitter of the solitary laser. Regions in white indicate a
timing jitter greater then $20$fs. In subplot (b) black marks the regions where the deterministic system has a non-periodic solution and
the semi-analytical method cannot be applied.
Parameters: $D=0.2$, $\alpha_g=2$, $\alpha_q=1.5$, others are as in table \ref{tab:Simparams}.
}\label{pl3}
\end{center}
\end{figure}

For the parameters used in Fig.~\ref{pl2} (b) the system is well behaved and the solutions are periodic, however
for other parameters, particularly for larger feedback strengths and non-zero amplitude-phase coupling,
this is not the case; quasi-periodic or chaotic dynamics can be observed. In such regions the semi-analytic approach
is invalid, however the timing jitter calculated by numerical methods is not meaningful in these non-periodic region either.
In Fig.~\ref{pl3} the timing jitter, calculated from the numerical (a) and semi-analytical (b) methods, is plotted in
dependence of $K$ and $\tau$ for $\alpha_g=2$ and $\alpha_q=1.5$. The timing jitter is given by the colour code, where blue regions
indicate a reduction in the timing jitter with respect to the solitary laser, red tones indicate an increase and white regions indicate a
timing jitter greater than $20$fs, indicative of a non-periodic pulse stream. In the black regions in Fig.~\ref{pl3} (b)
the solutions of the DDE system are non-periodic and the semi-analytic method is not applied. Good
agreement is observed between these two methods over
most of the parameter range depicted. The non-periodic regions indicated in subplot (b) coincide
with the very high timing jitter estimations obtained using the
numerical method.

A key difference between the two methods is that the semi-analytic method is based on the
numerical simulation of deterministic equations, while the purely numerical method requires
integration of a system of stochastic DDEs. Using the latter method one can run into problems that
arise due to the multiplicity of stable solutions found in this system. Since timing jitter estimation
requires averaging over many noise realisations, depending on the particular realisation, due to transient effects, the system
can land on different solutions. As different ML solutions can have slightly different inter-spike
interval times, the fully numerical estimation of the timing jitter can lead to erroneously large
values in such case \cite{SIM14}. This makes it difficult to perform
timing jitter calculations over a large parameter domain, as it is not easy to distinguish between
the above mentioned effect and a destabilisation of the pulse stream due to the feedback conditions.
Note that this is a different effect to switching between solutions within one time series. Such
difficulties are eliminated when using the
semi-analytic method, as in this case the estimation of the variance is based on the integration of
deterministic equations. Therefore, there are two main advantages to using the semi-analytic method
to calculate the timing jitter, compared with brute force methods involving numerical integration of
stochastic differential equations. Firstly, the aforementioned difficulties can be avoided, and
secondly, the computation times can be greatly reduced (by over a factor of 100) as averaging over many noise realisation is
not needed. This means that it can become feasible to calculate the timing jitter for longer
feedback delay times, which is of interest
due to the improved timing jitter reduction predicted for increased delay times \cite{OTT14} and for
better comparison with experiments, where typically very long feedback cavities are used
\cite{ARS13,LIN10e}.

\subsection{Delay length dependence of timing jitter}

\begin{figure}[h]
\begin{center}
\includegraphics[width=0.49\textwidth]{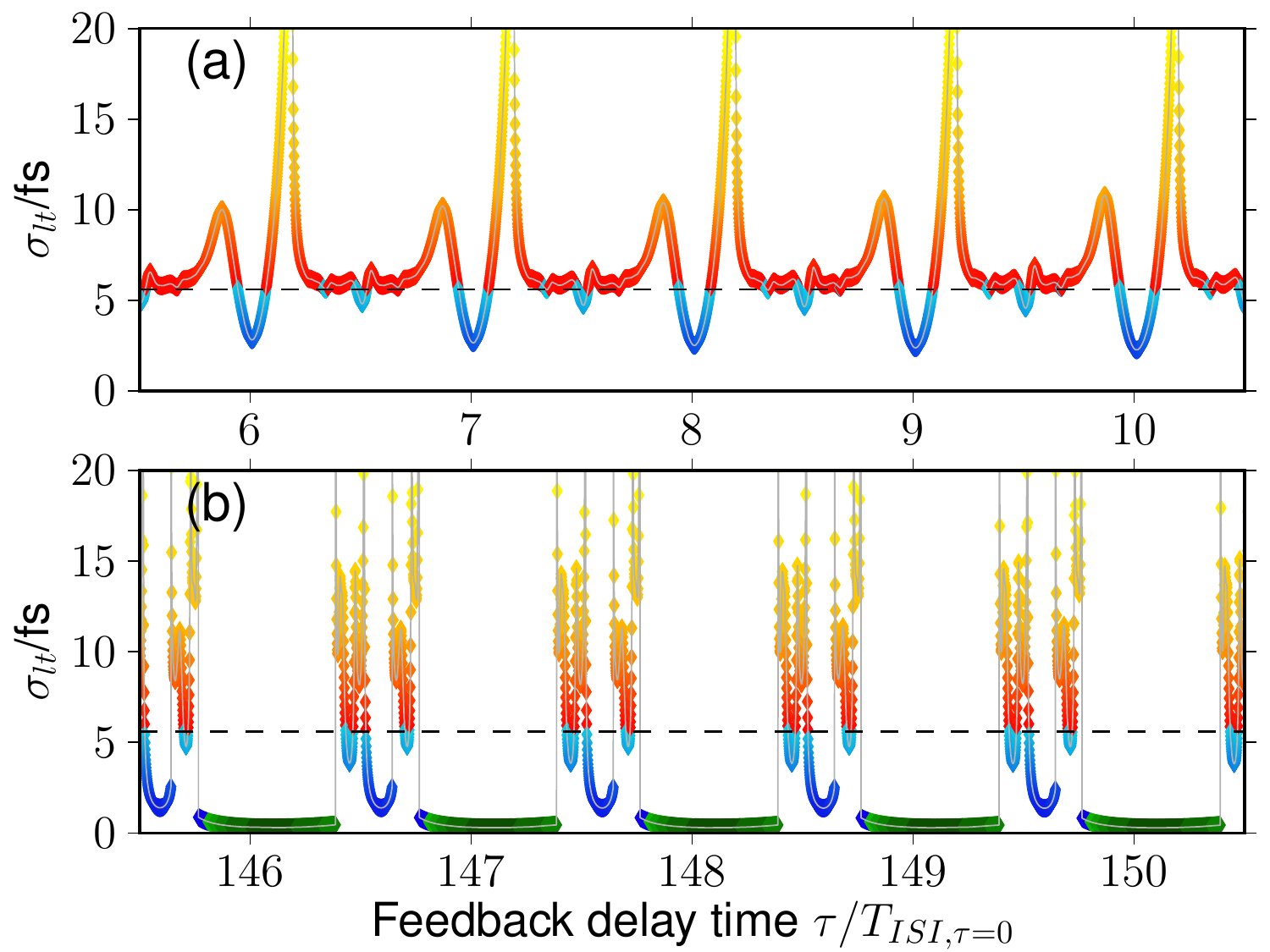}  \includegraphics[width=0.49\textwidth]{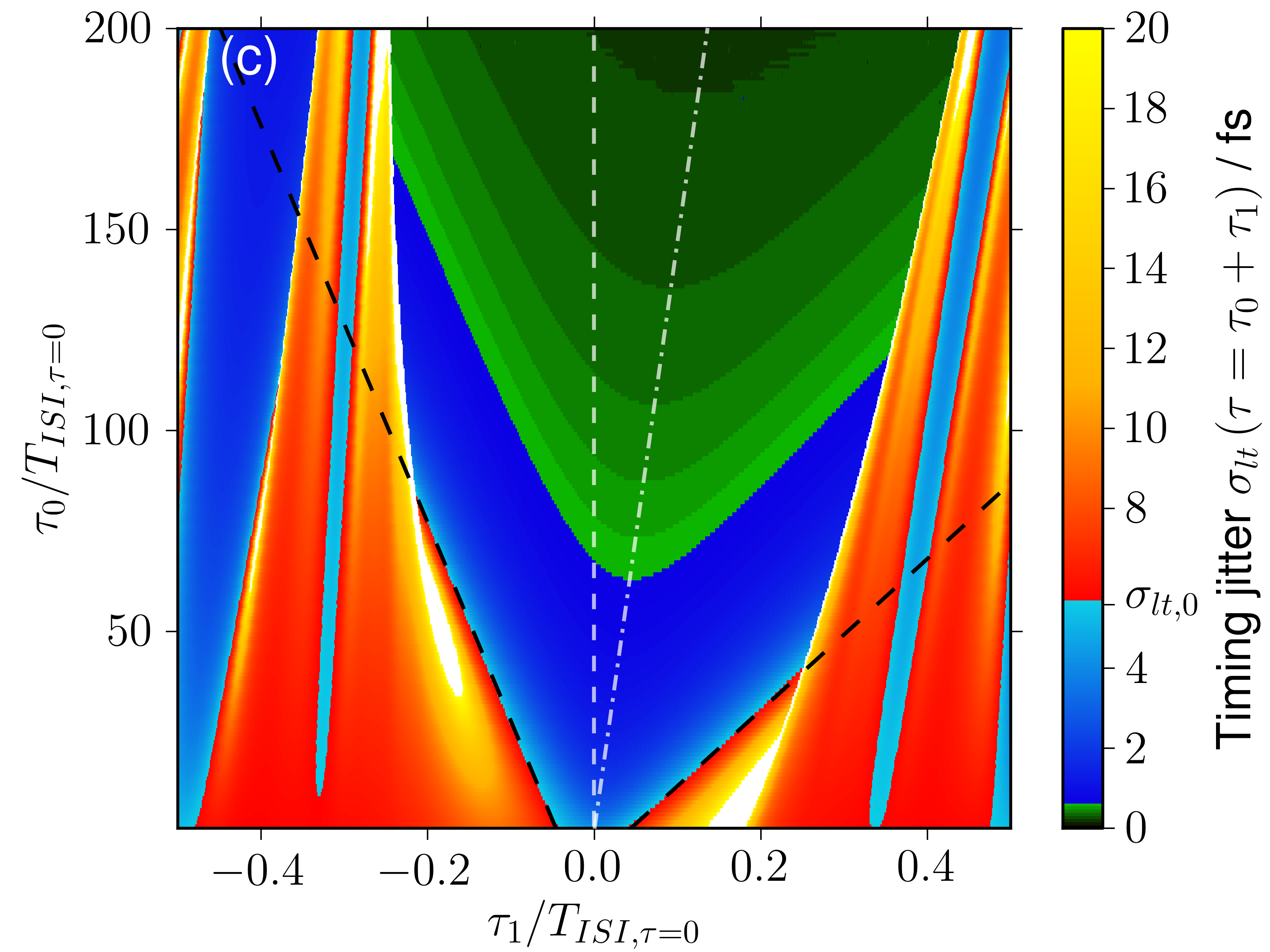}
\caption{(a) and (b) Timing jitter $\sigma_{lt}$ in dependence of the feedback cavity delay time.
The colour code indicates the
timing jitter according to the colour bar given in subplot (c). The black dashed line indicates the
timing jitter of
the solitary laser. (c) Timing jitter $\sigma_{lt}$ in dependence of the feedback cavity delay time,
where $\tau=\tau_0+\tau_1$ for any given point. The horizontal axis spans one $T_{ISI,\tau=0}$ and
is centered on an exact main resonance. The vertical axis indicates the number of the main
resonance. The timing jitter is indicated by the colour code and $\sigma_{lt,0}$ is the
timing jitter of the solitary laser.
Parameters: $K=0.1$, $D=0.2$, $\alpha_g=0$, $\alpha_q=0$, others as in table \ref{tab:Simparams}.
}\label{pl4}
\end{center}
\end{figure}

We now use the semi-analytic method to investigate how the timing jitter decreases with increased
resonant feedback delay times and how the width of the
frequency-pulling regions is affected by this
increase. In Fig.~\ref{pl4} the timing jitter is plotted as a function of $\tau$ in subplots (a)
and (b) for a short and a long $\tau$ range, respectively. The black dashed line indicates the
timing jitter of the solitary laser. The delay times are plotted in
units of $T_{ISI,\tau=0}$, the inter-spike interval time for zero delay
feedback (instantaneous feedback, $\tau=0$ and $K\neq0$), meaning that the resonant feedback
occurs at the integer delay values. ($T_{ISI,\tau=0}$ and $T_{ISI,0}$ only differ slightly. Here we choose $T_{ISI,\tau=0}$
as our reference because the period is the same for all $\tau=qT_{ISI,\tau=0}$, $q\in\mathbb{N}$, and we will
use this property in subsequent calculations.) In both (a) and (b) a timing jitter reduction is
observed for resonant feedback. For the longer delay times depicted in subplot (b) the timing jitter
reduction is greater and the frequency-pulling region
about the main resonances is wider. Changes in the frequency-pulling regions are not discernible over small $\tau$
ranges. To show the change
in dependence of $\tau$ more clearly a map of the timing jitter is shown in a $\tau-\tau$ plot in
subplot (c). In this plot both axes are related
to the delay time, the $\tau_1$ axis shows changes over one
$T_{ISI,\tau=0}$-interval , whereas the $\tau_0$ axis shows changes from one resonance to
the next. For
each point on this map the feedback delay time is given by $\tau=\tau_0+\tau_1$. The $\tau_1$ axis
is centered on the exact main resonances $\tau=qT_{ISI,\tau=0}$
and the $\tau_0$ axis gives the number $q$ of the main resonances. The timing jitter is
given by the colour code. Regions in blue and green indicate
a reduction in the timing jitter with respect to the solitary laser ($K=0$) and regions in red indicate an
increase in the timing jitter. In the green
regions the timing jitter is reduced by a factor of 10 or greater. For all $q$ values a reduction in
the timing jitter is achieved at the exact main resonances and for increasing $q$ the decrease in
the timing jitter can clearly be seen. It is seen from Fig.~\ref{pl4}(c) that for short
delays the width of the frequency-pulling regions, with reduced timing jitter, increases
approximately linearly with the number $q$. The edges of the frequency pulling region are marked by the
dashed black lines. At about $q=50$ the frequency-pulling region is
intersected by the solutions that correspond to higher order resonances ($p\tau=qT_{ISI,\tau=0}$, where $p=2,3,4,...$). This is due to a bistability
between the main and higher order resonant solutions \cite{OTT12a}. For the results
presented in subplot (c) of Fig.~\ref{pl4}, the same
initial conditions were used in the numerical simulations for all delay values. By performing a
sweep in $\tau$ (using the previous $\tau$ solution as the initial conditions for the next $\tau$) one can stay on the main resonant solution
in the bistable regions.

\begin{figure}[h]
\begin{center}
\includegraphics[width=0.6\textwidth]{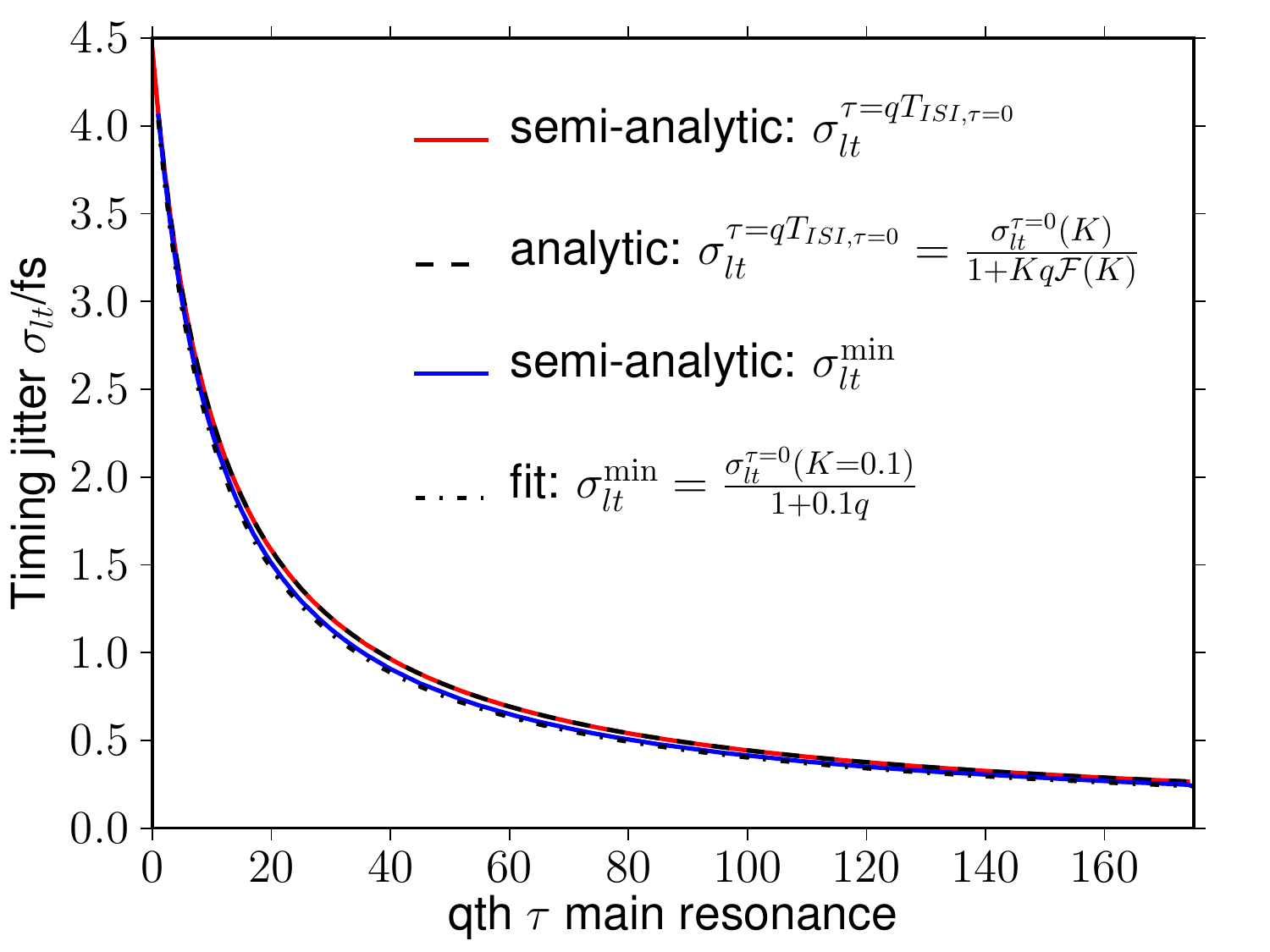}
\caption{Timing jitter $\sigma_{lt}$ at the exact main resonances (red solid line) and the minimum timing jitter in each
resonance region (blue solid line) as a function of the number $q$ of
the main resonance, calculated using the semi-analytic
method. The dashed line shows the timing jitter at the exact main resonances given by the analytic expression Eq.~\eqref{sigmares}. The dot-dashed
line shows the fit of Eq.~\eqref{sigmin} to the minimum timing jitter in each
resonance regions.
Parameters: $K=0.1$, $D=0.2$, $\alpha_g=0$, $\alpha_q=0$, others as in table \ref{tab:Simparams}.
}\label{pl5}
\end{center}
\end{figure}

In order to quantify the decrease in the timing jitter with increasing number $q$, we
have plotted the timing jitter at the main resonances in Fig.~\ref{pl5}. The red line shows the
results of the semi-analytic method for the exact main resonances $\tau=qT_{ISI,\tau=0}$ ($\tau$ values corresponding to the white dashed line in Fig.~\ref{pl4} (b)) and the
blue line shows the results of the semi-analytic method for the minimum timing jitter in each main
resonance frequency-pulling region ($\tau$ values corresponding to the white dot-dashed line in Fig.~\ref{pl4} (b)). The expression for the timing jitter at the main resonances, $\tau=qT_{ISI,\tau=0}$, can be
derived analytically using Eq.~\eqref{eq:jitter} and the bilinear form \eqref{eq:form}. At the
exact main resonances the solutions to Eqs.~\eqref{subeq:E}-\eqref{subeq:Q} are identical for all
$q$, and
the periodicity is the same as that of the laser with zero delay (instantaneous) feedback $T_0$ ($T_{ISI,\tau=0}$).
Therefore, for $\tau=qT_{ISI,\tau=0}$, Eq.~\eqref{eq:form} can be expressed as
\begin{gather}
\left[ \delta \psi^\dagger, \delta \psi \right](t) = \delta \psi^\dagger(t) \delta \psi(t)+\int_{-T}^{0}
 \dpsi^\dagger\left(t+r+T\right)B_0\left(t+r\right)\dpsi\left(t+r\right)dr\nonumber\\
  +K\int_{-T}^{0}
 \dpsi^\dagger\left(t+r+T\right)B_1\left(t+r\right)\dpsi\left(t+r\right)dr \nonumber \\
 +K\int_{-T-qT_{ISI,\tau=0}}^{-T}
 \dpsi^\dagger\left(t+r+T\right)B_1\left(t+r\right)\dpsi\left(t+r\right)dr. \label{bilin2}
\end{gather}
The last term on the right-hand side can be further simplified due to the time shift invariance and periodicity of the integrand, giving
\begin{equation}
\left[ \delta \psi^\dagger, \delta \psi \right]=\left[ \delta \psi^\dagger, \delta \psi \right]^{\tau=0} +Kq\int_{-T_{ISI,\tau=0}}^{0}
 \left(\dpsi^\dagger\left(t+r+T\right)\rb^{\mathrm{T}}B_1\left(t+r\right)\dpsi\left(t+r\right)dr, \label{bilin3}
\end{equation}
where the first three terms on the rhs of Eq.~\eqref{bilin2} are now expressed as $\left[ \delta \psi^\dagger, \delta \psi \right]^{\tau=0}$, which is the bilinear form for $\tau=0$ ($q=0$).
Equation~\eqref{eq:jitter} can thus be expressed as
\begin{equation}
 \sigma_{var}=\sqrt{D^2\int_{0}^{T_0}\left(\frac{\delta\psi_{0t,1}^{\dagger *}\left(t\right)}{\left[\dpsi_{1}^{\dagger *},\dpsi_{1}^*\right]^{\tau=0} +Kq \mathcal{F}'\lb K\rb}\rb^2+\left(\frac{\delta\psi_{0t,2}^{\dagger *}\left(t\right)}{\left[\dpsi_{1}^{\dagger *},\dpsi_{1}^*\right]^{\tau=0} +Kq \mathcal{F}'\lb K\rb}\rb^2 dt}, \label{svar2}
\end{equation}
where $\frac{\dpsi_{1}^{\lb \dagger \rb *}}{\left[ \dpsi_{1}^{\dagger *},\dpsi_{1}^*\right]}=\dpsi_{1}^{\lb \dagger \rb }$ and
\begin{equation}
 \mathcal{F}'\lb K\rb =\int_{-T_{ISI,\tau=0}}^{0}
 \delta\psi_{0t}^{\dagger *}\left(t+r+T\right)B_1\left(t+r\right)\dpsi_{1}^*\left(t+r\right)dr, \nonumber
\end{equation}
which is a function of $K$ but not of $\tau$. Finally, Eq.~\eqref{svar2} can be simplified to
\begin{equation}
 \sigma_{var}=\frac{1}{1+Kq\mathcal{F}' \lb K\rb}\sqrt{D^2\int_{0}^{T_0}\left(\delta\psi_{0t,1}^{\dagger \tau=0}\left(t\right)\rb^2+\left(\delta\psi_{0t,2}^{\dagger \tau=0}\left(t\right)\rb^2 dt}, \label{svar3}
\end{equation}
where $\delta\psi_{0t}^{\dagger \tau=0}=\lb \delta\psi_{0t,1}^{\dagger \tau=0},\delta\psi_{0t,2}^{\dagger \tau=0},\delta\psi_{0t,3}^{\dagger \tau=0},\delta\psi_{0t,4}^{\dagger \tau=0}\rb^\mathrm{T}$
is the solution fulfilling the biorthogonality condition for $\tau=0$ and $\mathcal{F}\lb K \rb =\frac{\mathcal{F}'\lb K\rb}{\left[\dpsi_{1}^{\dagger *},\dpsi_{1}^*\right]^{\tau=0}}$. The timing jitter for
resonant feedback, $\tau=qT_{ISI,\tau=0}$, is therefore given by
\begin{equation}
 \sigma_{lt}^{\tau=qT_{ISI,\tau=0}}= \frac{\sigma_{lt}^{\tau=0}\lb K\rb}{1+Kq\mathcal{F} \lb K\rb}, \label{sigmares}
\end{equation}
where $\sigma_{lt}^{\tau=0}\lb K\rb$ is the timing jitter for $\tau=0$. The curve obtained using this analytic expression is shown by dashed black line in Fig.~\ref{pl5}. A formula for the minimum jitter can not be derived in the same way as the inter-spike interval time changes with $q$. However,
fitting the minimum jitter curve for various feedback strengths we find that the relation
\begin{equation}
\sigma_{lt}^{\mathrm{min}}\approx\frac{\sigma_{lt}^{\tau=0}\lb K\rb}{1+Kq}, \label{sigmin}
\end{equation}
holds well for low feedback strengths. The fit is plotted in the black dot-dashed line in
Fig.~\ref{pl5}.

In the derivation of Eq.~\eqref{sigmares} contributions to the timing jitter from
eigenfunctions with negative eigenvalues, $\lambda<0$, are neglected. However, for
increased feedback delay lengths, the number of weakly stable Floquet multipliers close to
one increases. This leads to long transients in numerical simulations of the deterministic
system (Eqs.~\eqref{subeq:E}-\eqref{subeq:Q} $D=0$). These transient effects are accompanied
by fluctuations in the pulse heights, which have the periodicity of the feedback delay time.
Including noise in the system excites these transient amplitude fluctuations, which results in an
increased timing jitter, as, via the interaction with the gain and absorber media, changes in the
pulse height also lead to slight changes in the pulse positions.
Equation~\eqref{sigmares} is therefore only valid in the limit in which such effects can be neglected. For
the parameter values used in our simulations Eq.~\eqref{sigmares} holds for up
to $q\approx300$. These noise induced transient effects were observed
experimentally as side peaks in the phase noise spectra \cite{HAJ12,ARS13,DRZ13a}.

\section{Conclusions}

We have investigated the influence of optical feedback on the timing jitter of a passively ML semiconductor laser.
For resonant feedback we have derived an expression, Eq.~\eqref{sigmares}, for the analytical dependence of the timing jitter on the
feedback delay length, showing that the timing jitter drops off as approximately $1/\tau$ for $\tau\gg T$, as long as
amplitude jitter effects can be neglected. About the main resonant feedback delay lengths, frequency-pulling regions form,
in which the timing jitter is reduced with respect to the solitary laser. For small
feedback strengths $K$ the widths of these frequency-pulling regions
increase linearly with the number $q$ of the main resonance.
These results were obtained using a semi-analytical method, presented in this paper,
of calculating timing fluctuations in a DDE system describing the dynamics of a
passively ML semiconductor laser subject to optical feedback from an arbitrary number of feedback
cavities. The semi-analytical method shows good agreement
with methods based on direct numerical integration of the stochastic model and has the advantage of greatly reduced computation times.

\begin{acknowledgments}
The authors are thankful to D. Turaev for a very useful discission of the asymptotic analysis method.
L. Jaurigue thanks B. Lingnau for fruitful discussions and acknowledges support from the GRK
1558 funded by the DFG. A. Pimenov and A. G. Vladimirov acknowledge the support of SFB 787 of the
DFG, project B5. A. G. Vladimirov also acknowledges the support of 14-41-00044 of RSF at the Lobachevsky
University of Nizhny Novgorod. D. Rachinskii acknowledges the support of NSF through grant
DMS-1413223.
\end{acknowledgments}


\appendix
\section{Appendix: Derivation of the expression for the rate of the phase diffusion}

Here we derive formula \eqref{dbar} for the phase diffusion rate. Recall that $\psi_0(t)$ is a $T_0$-periodic ML solution of system \eqref{subeq:E}--\eqref{subeq:Q}.
Substituting the expression $\psi(t) = \psi_0(t) + \delta \psi(t)$ into this system, we obtain the linearized equations
\begin{equation}
 \frac{d}{dt}\delta\psi\lb t \rb = A\lb t\rb \dpsi\lb t\rb+\sum_{m=0}^{M}B_m\lb t-\tau_m'\rb \dpsi \lb t-\tau_m'\rb
 + D w(t),\label{linsys}
\end{equation}
where $A$ and $B_m$ are $T_0$-periodic Jacobi matrices of the linearization; $\tau_0'=T$, $\tau_m'=T+\tau_m$ for $m\ge 1$; and, $D w(t) = D (\xi_1(t), \xi_2(t), 0, 0)^{\mathrm{T}}$ is the small noise term.
Explicit expressions for the matrices $A(t)$ and $B(t)$ can be found in \cite{REB11}.
When there is no noise ($D=0$), the homogeneous system
\begin{equation}\label{eq:linearh}
-\frac{d}{dt}\delta\psi (t) + A(t) \delta \psi (t)  + \sum_{m=0}^M B_m(t - \tau_m') \delta \psi(t- \tau_m')= 0
\end{equation}
and its adjoint system, for a row vector $\delta \psi^\dagger(t) = (\delta \psi^\dagger_{1}, \delta \psi^\dagger_{2},\delta \psi^\dagger_{3}, \delta \psi^\dagger_{4})$,
\begin{equation}\label{eq:linearadj}
\frac{d}{dt}\delta \psi^\dagger(t) + \delta \psi^\dagger(t) A(t) + \sum_{m=0}^M \delta \psi^\dagger (t+\tau_m') B_m(t) = 0,
\end{equation}
have characteristic solutions (eigenmodes) of the form $\delta \psi(t) = \dpsi_{\lambda}(t) =
e^{\lambda t} p_{\lambda}(t)$ and
$\delta \psi^\dagger(t)=\dpsi_{\lambda}^{\dagger}(t) = e^{-\lambda t} p_{\lambda}^{\dagger}(t)$,
respectively, where functions $p_{\lambda}(t)$ and $p_{\lambda}^{\dagger}(t)$ are $T_0$-periodic and the complex value $\lambda$ is a Floquet exponent of \eqref{eq:linearh}.
The bilinear form \cite{HAL66, HAL77}
\begin{equation}\label{eq:form}
\left[ \delta \psi^\dagger, \delta \psi \right](t) = \delta \psi^\dagger(t) \delta \psi(t) +
\sum_{m=1}^M \int_{-\tau_m'}^0 \delta \psi^\dagger(t+r+\tau_m') B_m(t+r) \delta \psi(t+r) d r
\end{equation}
is instrumental in quantifying the effect of noise
along different eigendirections $\dpsi_{\lambda}(t)$ for the perturbed system \eqref{linsys},
because for every solution $\delta \psi(t)$ of \eqref{linsys} and every solution $\delta \psi^\dagger(t)$ of  \eqref{eq:linearadj}
the following relation holds at all times:
\begin{equation}\label{eq:motion0}
\frac{d [ \delta \psi^\dagger, \delta \psi](t)}{dt} =D \delta \psi^\dagger(t) w(t).
\end{equation}
Indeed,
\begin{gather}
 \frac{d}{dt} [\delta \psi^\dagger, \delta \psi](t) = \frac{d}{dt} \left(\delta \psi^\dagger(t)
\delta \psi(t) + \sum_m \int_{-\tau_m'}^0 \delta \psi^\dagger(s+t+\tau_m') B_m(s+t) \delta
\psi(s+t) ds\right)\notag\\
= \frac{d \delta \psi^\dagger(t)}{dt}  \delta \psi(t) + \delta \psi^\dagger(t)  \frac{d \delta
\psi(t)}{dt} + \frac{d}{dt}\sum_m \int_{t-\tau_m'}^t \delta \psi^\dagger(s+\tau_m') B_m(s) \delta
\psi(s) ds\notag\\
= -\left(\delta \psi^\dagger(t) A(t)  + \sum_m \delta \psi^\dagger(t+\tau_m') B_m(t)\right) \delta \psi(t)  \notag \\
+ \delta \psi^\dagger(t) \left(A(t) \delta \psi(t)   + \sum_m  B_m(t-\tau_m') \delta \psi(t-\tau_m') + w(t)\right)\notag\\
 +\sum_m (\delta \psi^\dagger(t+\tau_m') B_m(t) \delta \psi(t) - \delta \psi^\dagger(t) B_m(t-\tau_m') \delta \psi(t-\tau_m')) \notag\\
=D\, \delta \psi^\dagger(t) w(t). \notag 
\end{gather}
In particular, for every pair of solutions of the homogeneous systems \eqref{eq:linearh} and \eqref{eq:linearadj} ($D=0$),
the form $[ \delta \psi^\dagger, \delta \psi](t)$ is independent of time.
Eq.~\eqref{eq:motion0} also ensures the biorthogonality property
\begin{equation}\label{bior}
[\dpsi^\dagger_{\lambda}, \dpsi_{\mu}](t)\equiv 0
\end{equation}
for any pair of eigenfunctions of problems \eqref{eq:linearh} and \eqref{eq:linearadj} with $\lambda\ne\mu$.
Furthermore, Eq.~\eqref{eq:motion0} implies that for any solution $\delta \psi(t)$
 of the inhomogeneous problem \eqref{linsys}, the projection $y_\lambda(t) = e^{\lambda t}
[\dpsi^\dagger_{\lambda}, \delta \psi](t)$ satisfies the equation
\begin{equation}\label{eq:motion}
 \frac{d y_\lambda(t)}{dt} = \lambda y_\lambda(t) + D p^\dagger_\lambda(t) w(t)
\end{equation}
with the Langevin term $w(t)$. For $\Re \lambda < 0$, this equation
defines an Ornstein-Uhlenbeck type process with a uniformly bounded variance of order $D^2$.
On the other hand, for $\lambda = 0$, we obtain a process similar to the Brownian motion
with the variance that grows linearly with time as $D^2 t$. Hence, noise mostly affects
the projections of a solution of \eqref{linsys} onto
the neutral eigenmodes \eqref{eq:neutral} that have $\lambda=0$.
The two corresponding adjoint neutral eigenfunctions (that is, $T_0$-periodic solutions of the 
adjoint system \eqref{eq:linearadj}) can be normalized in such a way as to satisfy the relations
\begin{equation}\label{eq:orth}
\left[\delta \psi_{1} ^{\dagger},  \delta \psi_{1}  \right](t) = \left[\delta \psi_{2} ^{\dagger},  \delta \psi_{2}  \right](t)\equiv 1,
\qquad \left[\delta \psi_{1} ^{\dagger},  \delta \psi_{2}  \right](t) = \left[\delta \psi_{2} ^{\dagger},  \delta \psi_{1}  \right](t)\equiv 0.
\end{equation}
For stable mode-locked solutions $\psi_0(t)$ all the non-zero Floquet exponents of the linearized system have negative real parts.

Using the linearization, we can approximate the asymptotic phase of a solution to the nonlinear system \eqref{subeq:E}-\eqref{subeq:Q} by the formulas
\begin{equation}\label{phase}
\left[\delta\psi_{1}^{\dagger}, \Gamma_{-\varphi} \psi(t-\theta)-\psi_0\right](t+\theta)=
\left[\delta\psi_{2}^{\dagger}, \Gamma_{-\varphi} \psi(t-\theta)-\psi_0\right](t+\theta)=0,
\end{equation}
These equations define the ``time'' phase $\theta$ and the ``angular'' phase $\varphi$ implicitly for any given state $\psi(t+r)$ ($r\in [-\tau_M',0]$) of the nonlinear system.
Geometrically, \eqref{phase} is a codimension 2 linear subspace which is tangent to the surface
of constant asymptotic phases $\theta$, $\varphi$ at the point where this surface intersects the torus of shifted periodic solutions $\Gamma_\varphi\cdot \psi_0(t+\theta)$
in the state space of the system. As we consider solutions that remain within a small distance of order $D$ from this torus,
the error between the asymptotic phase and its approximation \eqref{phase}
is of next order $D^2$. Also, note that Eqs.~\eqref{phase} themselves can be used as an alternative definition of the phase,
because these equations define a foliation of a small tubular neighborhood surrounding the torus of periodic solutions
by non-intersecting surfaces $\theta=const,\,\varphi=const$.

In order to derive the equation for the evolution of the phase, we differentiate Eqs.~\eqref{phase} with respect to $t,\theta,\varphi$.
Using symmetry, one obtains from Eq.~\eqref{eq:motion0} 
the relationship
\begin{equation}\label{d1}
\frac{\partial}{\partial t} \left[\delta\psi_{i}^{\dagger}, \Gamma_{-\varphi}
\psi(t-\theta)-\psi_0\right](t+\theta)=D\, \delta \psi_{i}^\dagger(t+\theta) \Gamma_{-\varphi} w(t)
\end{equation}
for $i=1,2$. When differentiating the bilinear form $\left[\delta\psi_{i}^{\dagger}, \Gamma_{-\varphi}
\psi(t-\theta)-\psi_0\right](t+\theta)$ with respect to $\theta$ and $\varphi$,
we omit the terms that are proportional to $\psi -\Gamma_\varphi 
\psi_0(t-\theta)$, because these therms have the order $D$ in the small vicinity of the cycle that we consider.
In this approximation, we obtain
 \begin{equation}\label{d2}
\frac{\partial}{\partial \theta} \left[\delta\psi_{i}^{\dagger}, \Gamma_{-\varphi} \psi(t-\theta)-\psi_0\right](t+\theta)=-\left[ \delta\psi_{i}^{\dagger}, \delta \psi_{1} \right](t+\theta),
\end{equation}
 \begin{equation}\label{d3}
\frac{\partial}{\partial \varphi} \left[\delta\psi_{i}^{\dagger}, \Gamma_{-\varphi} \psi(t-\theta)-\psi_0\right](t+\theta)=-\left[ \delta\psi_{i}^{\dagger}, \delta \psi_{2} \right](t+\theta).
\end{equation}
Combining relationships \eqref{eq:orth}--\eqref{d3},
we arrive at 
the coupled system of stochastic equations \eqref{rate} that describe the slow evolution of the variables $\theta$ and $\varphi$.

Finally, using the Feynman-Kac formula, we obtain the Fokker-Planck equation for the joint probability density
$p(t,\theta,\varphi)$ of the stochastic process \eqref{rate}:
\begin{equation}\label{fp}
\frac{\partial p}{\partial t}=\left(\frac12 \frac{\partial^2}{\partial \theta^2} (d_{11}\, p )
+ \frac{\partial^2}{\partial \theta\partial\varphi} (d_{12}\, p)
+ \frac12\frac{\partial^2}{\partial \varphi^2}(d_{22}\, p)\right).
\end{equation}
This equation has variable diffusion coefficients
$$
\begin{array}{ll}
d_{11}=D^2\bigl(\bigl(\delta \psi_{1, 1}^\dagger\bigr)^2+\bigl( \delta \psi_{1, 2}^\dagger\bigr)^2\bigr)(t+\theta),\\
d_{22}=D^2\bigl(\bigl(\delta \psi_{2, 1}^\dagger\bigr)^2+\bigl( \delta \psi_{2, 2}^\dagger\bigr)^2\bigr)(t+\theta),\\
d_{12}=D^2\bigl(\delta \psi_{1, 1}^\dagger \delta \psi_{2, 1}^\dagger + \delta \psi_{1, 2}^\dagger \delta \psi_{2, 2}^\dagger\bigr)(t+\theta),
\end{array}
$$
where $\delta\psi_{i, k}^\dagger$ are the coordinates of the 4-dimensional vector-functions $\delta\psi_{i}^\dagger$.
Since, for $D\ll 1$, the probability density changes slowly, Eq.~\eqref{fp} can be averaged over the period $T_0$ of the functions $d_{ij}(t+\theta)$,
resulting in the diffusion equation with constant coefficients $\bar d_{ij}$ (see, for example, \cite{averaging}). The averaged coefficient $\bar d_{11}$
that approximates the rate of diffusion of the phase $\theta$ is defined by  formula \eqref{dbar}. 

\bibliographystyle{prwithtitle_aglabel.bst}


\end{document}